\documentclass[12pt]{article}

\usepackage{amstext,amsmath,amssymb,amsfonts}
\usepackage{bbm}
\usepackage[latin1]{inputenc}
\usepackage{hyperref}
\usepackage{amsthm}
\usepackage{color}
\usepackage{graphicx}
\usepackage{geometry}
 
\newtheorem{definition}{Definition}
\newtheorem{theorem}{Theorem}

\newtheorem{proposition}{Proposition}
\newcommand{\hX}{\hat{X}}
\newcommand{\hY}{\hat{Y}}
\newcommand{\Tr}{\mathrm{Tr}}
 
\newcommand{\be}{\begin{equation}}
\newcommand{\ee}{\end{equation}}
  
\newcommand{\bea}{\begin{eqnarray}}
\newcommand{\eea}{\end{eqnarray}}

\begin{document}
\begin{flushright}
 
\end{flushright}

\vspace{20pt}

\begin{center}

{\Large\bf 
A Givental-like Formula and \\
\medskip
Bilinear Identities for Tensor Models.}
\vspace{20pt}

St\'ephane Dartois${}^{a,b}$
\footnote{e-mail: stephane.dartois@lipn.univ-paris13.fr or stephane.dartois@outlook.com},

\vspace{10pt}

\begin{abstract}
\noindent In this paper we express some simple random tensor models in a Givental-like fashion \textsl{i.e.} as differential operators acting on a product of generic 1-Hermitian matrix models.  
Finally we derive Hirota's equations for these tensor models. Our decomposition is a first step towards integrability of such models. 
\end{abstract}
\end{center}


\setcounter{footnote}{0}
\setcounter{lemma}{0}
\setcounter{theorem}{0}

\section{Introduction}

Tensor models are the natural generalization of matrix models. Initially introduced in the context of nuclear physics, random matrices
have been applied extensively from statistical physics to number theory. A major development was the introduction of interacting
random matrix models and of their $1/N$ expansion. 
Indeed it was discovered that the Feynman graphs of these models implement consistently a sum over cellular decompositions 
of Riemann surfaces naturally encoding the topology of the discretized surfaces  \cite{'tHooft:1973jz}. 
This feature made them central both to string theory and to two dimensional quantum gravity.
Indeed the double scaling limit of matrix models provides a road towards the still ongoing non-perturbative definition of string theory. 
 Moreover the relationship between the continuum Liouville formulation and discretized matrix models plays an ever increasing 
role for the understanding of two-dimensional quantum gravity. 

\medskip
Early tensor models were introduced \cite{oldgft1,oldgft2} in order to generalize to higher dimensions this great success of matrix models.
Unfortunately they turn out to be difficult to handle analytically. Some key concepts of matrices (eigenvalues, characteristic polynomials, determinants)
simply do not generalize (in a easily computable way) to higher rank tensors.  Moreover the much richer geometries in higher dimensions obviously bring new challenges. Which description of geometry is 
most naturally and in the simplest way associated to a Feynman theory? This question, which was certainly not central from a purely mathematical point of view, 
is the crucial one for a field theoretic-type quantization of gravity in higher dimensions. Graph Encoded Manifolds (GEM) theory and crystallization \cite{lins, FG} in fact turned out to provide an answer to 
\emph{that} question. Indeed colored triangulations are dual to simple field theoretic combinatorics.

\medskip
Following this observation, \emph{colored} tensor models were introduced \cite{color}. Unexpectedly they also solve many associated difficulties.
First and foremost they allowed to find a tensor analog of the matrix $1/N$ expansion \cite{expansion3}. 
 
A major new feature of this expansion is that it is not of topological nature (at least not in a naive way). 
The full meaning of the parameter governing this expansion, 
called the \emph{degree}, is still unclear, although it can be computed rather easily as the sum of the genera of normal surfaces embedded in the cellular decomposition of the discretized geometry \cite{ryan}. Progress followed quickly, in particular through computation of single \cite{critical,KOR} and double scaling limits of such models \cite{Dartois:2013sra,GS,Bonzom:2014oua},
and the inclusion of matter in the corresponding random geometry \cite{IsingD,spinglass}.

\medskip
From the start tensor models were also related to the group field theory (GFT) approach to quantum gravity \cite{Oriti:2013aqa,Oriti:2013jga,Krajewski:2012aw}. 
This approach implements a sum over spin networks of loop quantum gravity \cite{LQG2} as a quantum field theory defined on a  Lie group. 
Tensor models provided GFT with a consistent class of interactions and observables which generalize the concept 
of locality to a background-independent formulation. This improvement allowed to renormalize GFT \cite{BenGeloun:2011rc,Carrozza:2013mna}. A complete study has been achieved in \cite{Carrozza:2014rba}, interesting physical perspectives are enumerated in \cite{Rivasseau:2013uca}.

\medskip
Over the years, matrix models have developed a lot of additional exciting features. First they could be used to tackle combinatorial problems,
such as counting a variety of $2$-dimensional maps \cite{Itzykson:1979fi}. Using a new class of matrix models, Kontsevich proved Witten's conjecture on
generating functions for intersection numbers of moduli spaces. Matrix models also have rich integrability
properties, unraveled through orthogonal polynomials, KdV hierarchy and Hirota's equations.
More recently Givental introduced a decomposition of solutions to the multi-components KP hierarchy using
matrix models \cite{Giv1, Giv2}. An other important development which aroses from matrix models around the same time 
is the topological recursion \cite{Eynard:2004mh}, which has grown into a polyvalent technique allowing to solve 
many problems of algebraic and enumerative geometry \cite{EOAlgRM}. The relation between these two techniques has been shown in \cite{DBOSS}.

\medskip
It is then natural to ask whether tensor models inherit, at least partly, of these important mathematical features.  
In this paper we derive a decomposition formula that looks like (but is not) a Givental decomposition formula for the 
simplest random tensor model, namely the quartic melonic model (at any rank). In fact we describe the partition function of this 
model as the action of a differential operator on a product of Hermitian $1$-matrix models (the number of matrix models being given by the rank of the tensor models). We end this paper by deriving bilinear identities for the tensor model by deforming the Hirota's equations satisfied by the Hermitian $1$-matrix model. 

\medskip
This paper is structured as follow:
\begin{itemize}
 \item Section \ref{sec:Matrix} about matrix models recalls the well known results on the subject .
 \item Section \ref{sec:color} introduces tensor models more formally, then specializes to the quartic melonic tensor model and derives the intermediate field representation for it.
 \item Section \ref{sec:Decomp} shows the decomposition formula for tensor model into Hermitian matrix models.
 \item Section \ref{sec:bilin} recalls the basic Hirota's equations for matrix models and derives bilinear identities for the quartic melonic tensor model from them.
\end{itemize}

\section{Matrix Models}

\label{sec:Matrix}

\subsection{Generalities on matrix models.}
For pedagogical reasons I present some known results about matrix models in this section, this should allow readers coming from different communities to read this paper easily. Most of the material can be found in \cite{matrix, Mehta}.\newline

{\bf Matrix $1/N$ development:} \newline
 We recall briefly the $1/N$ development of matrix models. Consider the matrix model defined by
 \be
  Z[t_4,N]=\int dM \exp\left(-N(\frac{1}{2}\Tr(M^2) +\frac{t_4}{4} \Tr(M^4))\right),
 \ee
 with $N$ the size of the matrix. At the formal level this is a generating function for quadrangulations. The free energy $F=\ln Z$ expands as $F=\sum_{g\ge 0} N^{2-2g} F_g(t_4)$ where the $F_g$'s are generating functions of quadrangulations of genus $g$ for the counting variable of quadrangles $t_4$. In the limit $N \rightarrow \infty$ only the leading order survives {\it i.e.} the term $F_0$ counting the planar quadrangulations, so to say quadrangulations of the sphere $S^2$.
 One can compute the two points function $G_2(t_4)= \frac{1}{N} \langle \Tr(M^2)\rangle$ in this limit and recover the Tutte's result \cite{Tutte} for planar rooted quadrangulations:
 \be 
  G_2(t_4,N=\infty)= \sum_n 2 \frac{3^n}{(n+2)(n+1)} \binom{2n}{n} t_4^n.
 \ee
 
{\bf Matrix double scaling limit:}

One expands the free energy in $N$:
\be 
F(t_4, N)=\sum_{g\ge 0} N^{2-2g} F_g(t_4).
\ee
All $F_g$'s have a critical point at $t_4=t_c$. Roughly they behave as $F_g \sim C_g(t_4 - t_c)^{\frac{5}{4}(2-2g)}$.
By setting $x=constant= N(t_4-t_c)^{5/4}$ while $N\rightarrow \infty$ and $t_4 \rightarrow t_c$ one gets the double scaling limit of $F$:
\be
F(x)=\sum_{g\ge 0} x^{2-2g} C_g.
\ee
The $C_g$ coincide with the correlations functions of Liouville gravity. This corresponds to the continuum limit of matrix models. One can thus understand that $F(x)$ should satisfy some differential equations, that is, the differential equation satisfied by the Liouville partition function. This equation is of the Painlev\'e type. \newline

{\bf Orthogonal polynomials and integrability: }

This presentation is based on \cite{Itzykson:1979fi,DiFra}. Orthogonal polynomials allows to compute exactly the partition function of matrix model. For the sake of definitness consider the matrix model defined above with potential $V(M)= \frac{t_4}{4} \Tr{M^4}$.
The orthogonal polynomials are the only polynomials which are monic and orthogonal with respect to the measure $\exp(-\frac{t_4}{4}x^4)dx$:
\be
\int P_{N,t_4}(x)P_{M,t_4}(x)\exp(-N\frac{t_4}{4}x^4)dx = \delta_{N,M} K_N
\ee
$K_N$ a proportionality factor. This can be used to solve matrix models because the orthogonality relations determine completely the model. For instance an orthogonal polynomial  
is provided by $P_{N,t_4}(x) = \langle \det(x-M) \rangle$. Following \cite{Itzykson:1979fi}, changing variables to eigenvalues leads to:
\bea
P_{N,t_4}(x) &=& \langle \det(x-M) \rangle \nonumber \\
&=& \frac{1}{Z_N}\int \prod_{i=1}^N d\mu(x_i)(x-x_i)\prod_{i<j}(x_i-x_j)^2,
\eea
where $d\mu(x) = \exp(-NV(x)) dx$. Then write: 
\bea
Z_N \int d\mu(x) P_{N, t_4}(x) x^M = \int d\mu(x) \int \prod_{i=1}^N d\mu(x_i) (x-x_i) \Delta(\{x_j\})^2 x^M \nonumber \\
=\int \prod_{i=1}^{N+1} d\mu(x_i)  \Delta(\{x_j\}_{j=1\cdots N+1})\Delta(\{x_j\}_{j=1\dots N}) x_{N+1}^M\nonumber \\
= \frac{1}{N+1}\sum_k(-1)^{N+1-k}\int \prod_{i=1}^{N+1} d\mu(x_i) \Delta(\{x_j\}_{j=1\cdots N+1})\Delta(x_1,\cdots,\hat{x_k},\cdots , x_{N+1} ) x_{k}^M  
\eea
Noticing that: 
\bea
\frac{1}{N+1}\sum_k(-1)^{N+1-k} \Delta(x_1,\cdots,\hat{x_k},\cdots , x_{N+1} ) x_{k}^M = \begin{vmatrix}
1 & x_1 &\cdots & x_1^{N-1} & x_1^M \\
1 & x_2 & \cdots & x_2^{N-1} & x_2^M \\
\vdots & \cdots & \cdots & \cdots & \vdots \\
1 & x_{N+1} & \cdots & x_{N+1}^{N-1} & x^M_{N+1}  
\end{vmatrix}
\eea 
this determinant vanishes for $M\le N-1$. For $M=N$:
\bea 
Z_N \int d\mu(x) P_{N, t_4}(x) x^N =\frac{1}{N+1}\int \prod_{i=1}^{N+1} d\mu(x_i)  \Delta(\{x_j\}_{j=1\cdots N+1})^2 = \frac{Z_{N+1}}{N+1} 
\eea
This implies $K_N = \frac{Z_{N+1}}{(N+1) Z_N}$. So one computes $K_N$ as a function of the $Z_N$'s:
\bea
K_N= \frac{Z_{N+1}[t_4]}{(N+1) Z_N[t_4]}.
\eea
Finally this leads to $Z_N[t_4]=N!  \prod_{i=1}^{N-1}K_i$. One then derives recursion relations for the $K_N$ (for instance see \cite{DiFra} for a general description of these problems).

\subsection{Generic matrix model and Kontsevitch model.}
In this section we introduce the generic matrix model.
\begin{definition}
We define the Hermitian one matrix model by the partition function:
\begin{equation}
Z_{1MM}[\{t_p\}_{p=0...\infty},N]= \int_{H_N} dM \exp(-\frac{N}{2}\Tr(M^2) - N\sum_{p\ge0} t_p \Tr(M^p))
\end{equation}
$H_N$ being the space of $N\times N$ hermitian matrices. This partition function has to be understood at the formal level.
\end{definition}
The second term entering the definition of the generic matrix model is called the generic potential, each term of its development is an invariant of the matrix $M$ in such a way that the "action" is univalued. For these invariants to be independent one has to take the limit of big size $N$ of the matrix. Each invariant can be labelled by an integer $p$, and we introduce one coupling constant for each of these invariants.
We consider these coupling constants as formal parameters of a formal series obtained by expanding the exponential and interchanging the order of summation. The integral representation introduced above is just a reminder for writing the term of the corresponding formal series (although this integral is well defined for negative values of the coupling constant).  

The Kontsevitch model is the model computing the intersection numbers of moduli spaces of Riemann surfaces of genus $g$ and $n$ punctures.
\begin{definition}
The Kontsevitch model is here defined by :
\begin{equation}
 Z_K[\Lambda]= \int_{H_N} dX \exp\Bigl(-\Tr(X\Lambda X)+ i\Tr(X^3)\Bigr)
\end{equation}
where $\Lambda$ is a diagonal matrix. We call the Miwa's coordinates the $T_k = \frac{1}{k} \Tr(\Lambda^k)$.
\end{definition} 

These two models are important since they are used to decompose solutions of the multi-component KP hierarchy as an intertwining operator acting on a product of the matrix models described above \textsl{i.e.} prototypically:
\be
\mathcal{Z}= e^{\mathcal{U}}\prod_k Z_{1MM}[\{t^k_p\}]=e^{\mathcal{\hat{U}}} \prod_k Z_K[\Lambda_k],
\ee
where $\mathcal{Z}$ is  given from a spectral curve $S(x,y)$ by $\mathcal{Z}=e^{-F(S)}$. $F(S):=\sum_{g\ge 0} N^{2-2g}F_g(S)$ is the generating function of genus $g$ symplectic invariants\footnote{\textsl{i.e.} invariant through symplectomorphism of the spectral curve. This definition actually corresponds to a cohomological field theory.} (the $F_g$'s) of $S$. The intertwining operator $\mathcal{U}$ (resp. $\mathcal{\hat{U}}$) is a differential operator quadratic in the $t^k_p$'s and $\frac{\partial}{\partial t^k_p}$ (resp. the $T_p^k$'s and $\frac{\partial}{\partial T_p^k}$).

In order to get a glimpse of the difference between these two decompositions we give some more details (everything and much more can be found in the litterature). A spectral curve $S$ is a compact Riemman surface $\Xi$  endowed with two meromorphic functions\footnote{In the case of meromorphic functions  the spectral curve is said to be algebraic. Moreover if $\Xi$ is of genus $0$ it is rationnal.} $x,y \in \mathcal{M}(\Xi)$ 
satisfying an algebraic equation $S(x,y)=0$\footnote{Exists because of the algebraic character of the spectral curve.}. One defines the 1-form $\omega_1^0 = y dx$ on $\Xi$ and $\omega^0_2 = B$ is the Bergmann kernel of $\Xi$ (we need a choice of polarization of $\Xi$ to define everything properly). Denotes the poles of $\omega^0_1$ by $\{\alpha_i\}$ and its branch points by $\{a_i\}$.
The operator $U$ can be computed by decomposing the global Virasoro constraints (equivalently the "loop equations") locally on the poles of $\omega_1^0$ leading to the results of \cite{Orantin(2008)}, on the other hand this decomposition can be performed locally around the zeroes of $\omega^0_1$, this leads to the expression of $\hat{U}$ used to decompose on Kontsevitch tau functions.
More formally the loop equations can be rewritten as:
\begin{equation}
\mathcal{L}(p)\mathcal{Z}=0,\text{ } \forall p \in \mathcal{C}.
\end{equation}
The construction of the $\mathcal{L}(p)$ being ensured by the data given above. In the limit $p\rightarrow \alpha_i$ these operators projects onto local operators that can be described as:
\begin{equation}
\mathcal{L}(p)\sim \sum_{n=1}^{\infty} \frac{dz_i(p)}{z_i(p)^{n+1}}L_n^i.
\end{equation}
The $z_i$'s being local coordinates around the poles $\alpha_i$. With the Virasoro operators $L_n^i$ taking the usual form for Hermitian 1-matrix models: 
\begin{equation}
L_n^i=\frac{1}{N^2}\left( 2n \frac{\partial}{\partial t_n^i} +\sum_{k=1}^{n-1} \frac{\partial}{\partial t_{n-k}^i \partial t_{k}^i}\right) +\sum_{p=1}^{d_i}(p+n)t_n^i\frac{\partial}{\partial t_{n+p}^i}.
\end{equation}
Analogously the projection can be made onto local operators defined around the branch points $\{a_i\}$.
These operators are described in the formula (11) of \cite{matrixMtheory}.

\section{Tensor Models} \label{sec:color}

In this section we introduce briefly the general framework of tensor models. 
These models have been introduced in the $90$'s in order to mimic the success of matrix models in more than two dimensions, they have been constructed in order to give a definition of a 'sum over geometries' for three and more dimensions. Unfortunately they were at that time very difficult to handle analytically and the problem of generating well controlled triangulations was not understood \cite{lost}. 
Tensor models was then abandonned. Recently, Razvan Gurau revived interest in tensor models by constructing a (colored) model generating controlled triangulations, for which he was able to construct a $1/N$ expansion. The original point of view evolved after the 'uncoloring' paper \cite{uncoloring}.
For more details one can look in general references on the subject, for instance the necessary background is contained in \cite{uncoloring, review}.  

\subsection{Tensor invariants and generic $1$-tensor model.}

The uncolored point of view can be described as follows. The action of tensor model should be univalued when seen as a function on the vector space of tensor, thus it should be constructed out of tensor invariants, in fact this is how matrix models are constructed, the trace being the invariant.

First we shortly introduce the tensor invariants. To this aim we consider the tensors as multilinear forms on a direct product of vector spaces. A really nice and more detailed presentation of them is done in the second section of \cite{Delepouve:2014bma}. Consider a Hermitian space $(V, h)$ of complex dimension $\text{dim} V = N$, $h$ being the Hermitian product on $V$. $h$ induces the usual isomorphism $V\rightarrow V*$ by $v \mapsto v* = h(v,.)$.
Denoting $\{u_i\}_{i=1...N}$ a basis of $V$ and the dual basis $\{h(u_i,.)=\tilde{u}_i\}_{i=1...N}$, the coordinates of a vector in $V$ and its dual in $V*$ are related by complex conjugation from the property of the Hermitian product:
\be
v*=h(v,.)=h(\sum_i v_i u_i, .)= \sum_i \bar{v_i} \tilde{u}_i
\ee  

A rank $D$ tensor $T$ is a multilinear form $T: V^{\times D} \rightarrow \mathbb{C}$ one can write in a basis:
\be 
T=\sum_{\{i_p\}_{p=1...D}} T_{i_1...i_D} \tilde{u}_{i_1}\otimes...\otimes \tilde{u}_{i_D}.
\ee
Moreover the dual (denoted $\bar{T}$) of $T$ (\textit{i.e.} the multilinear form on $V*^{\times D}$) is written in the basis $\{u_i\}_{i=1...N}$: 
\be 
\bar{T}=\sum_{\{i_p\}_{p=1...D}} \bar{T}_{i_1...i_D} u_{i_1}\otimes...\otimes u_{i_D} 
\ee
by the property of the induced Hermitian product on a tensor product of Hermitian spaces.

By invariants of tensor we actually mean that the constructed quantity is invariant under any change of basis of $V$ and $V*$. If we change the basis by an element $g^{-1} \in GL(N,\mathbb{C})$ the coordinates of a vector $v$ are changed by the matrix $U(g)$ of $g$ and the coordinates of the dual vector are changed by $g^{-1}$. This induces the change of basis in the tensor product space, and thus on tensors:
\bea 
T_{i_1\cdots i_D}'= \sum_{j_1\cdots j_D} U(g_1^{-1})_{i_1 j_1} U(g_2^{-1})_{i_2 j_2} \cdots U(g_D^{-1})_{i_D j_D}T_{j_1\cdots j_D} \\
\bar{T}_{i_1...i_D}'= \sum_{j_1\cdots j_D} U(g_1)_{i_1 j_1} U(g_2)_{i_2 j_2} \cdots U(g_D)_{i_D j_D}\bar{T}_{i_1...i_D}
\eea      
This observation allows us to describe the possible tensor invariants. The invariants of order $2p$ are $p$-linear in both $T$ and $\bar{T}$. Using the tranformation rule given above, one notices that the only requirement for the quantity to be invariant is that the indices of a $T$ should contract to the indices of a $\bar{T}$ with respect to their position, the first index of a $T$ contracting with the first index of a $\bar{T}$ and so on. Thus to describe an invariant of order $2p$ one only has to describe the contraction pattern of the $p$ $T$'s with the $p$ $\bar{T}$'s. This can be represented by bipartite graphs with colored edges. With the convention that the $T$'s are represented by white vertices with $D$ half-edges indexed from $1$ to $D$ representing the position of the indices of $T$ and the $\bar{T}$'s by black vertices with $D$ half-edges also indexed from $1$ to $D$ representing the position for the $\bar{T}$. The contraction of the $j^{th}$ index of a $T$ with the $j^{th}$ of a $\bar{T}$ is then represented by contracting the respecting half-edges in the graph. Thus the set of invariants of order $2p$ are represented by all bipartite regular graphs of valence $D$ with a proper $D$-coloration of the edges.
\begin{figure}
\begin{center}
 \includegraphics[scale=1.2]{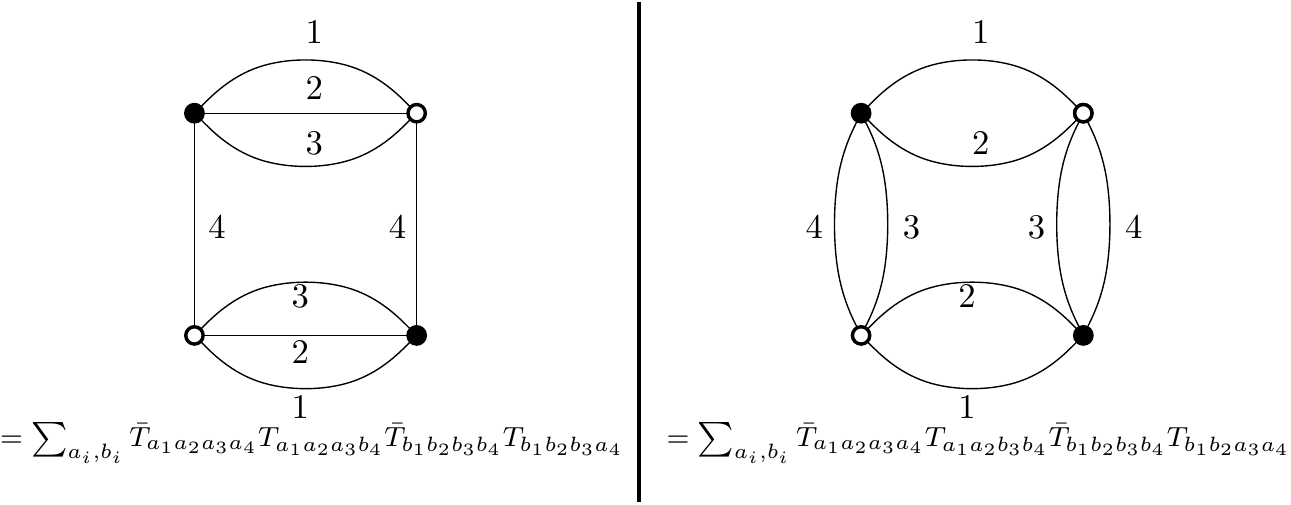}
 \caption{Two examples of tensor invariants for a rank $4$ tensor.}
\end{center}
\end{figure}
For any $D$-colored graph we define its jackets.
\begin{definition}
A colored jacket $\mathcal{J}$ is a ribbon graph associated to a $D$-colored graph $\mathcal{G}$ with 1-skeleton the graph $\mathcal{G}$ and faces made of graph cycles of colors $(\tau^q (0), \tau^{q+1} (0))$ for $\tau \in \mathcal{S}_D$ a cyclic permutation, modulo the orientation of the cycle (\text{i.e.} $\tau^{-1}$ leads to the same jacket).  
\end{definition}
From this we define the degree:
\begin{definition}
The \emph{degree} $\omega: \{D\mbox{-colored graphs}\}\rightarrow \mathbb{N}$ associates a positive integer to a $D$-colored graph $\mathcal{G}$ by:
\begin{equation}
\omega(\mathcal{G})=\sum_{\mathcal{J}(\mathcal{G})} g_{\mathcal{J}},
\end{equation}
{\it i.e.} it is the sum of the genera of all the jackets $\mathcal{J}$ of $\mathcal{G}$. 
\end{definition}
We are now ready to define the generic tensor model.
\begin{definition}
The generic tensor model of dimension $D+1$ is defined by the partition function
\bea
Z[N,\{t_{\mathcal{B}}\}]= \int dT d\bar{T} \exp\Bigl(-N^{D-1} \sum_{\mathcal{B}}N^{-\frac{2}{(D-2)!}\omega(\mathcal{B})} t_{\mathcal{B}}\mathcal{B}(T, \bar{T})\Bigr),
\eea
where $\mathcal{B}$ runs over the regular $D$-colored graphs indexing the invariants. The $t_{\mathcal{B}}$ are the coupling constant, the one corresponding to the only invariant of order $2$ often being fixed to one.
$\mathcal{B}(\cdot,\cdot)$ being the invariant of $T$ and $\bar{T}$ indexed by the graph $\mathcal{B}$. $\omega(\mathcal{B})$ is the degree of $\mathcal{B}$. 
\end{definition}
One notices that the jackets are specifics. In fact in the $4$-colored graphs case they provide cellular decompositions of Heegaard surfaces\footnote{If one wants to be precise, one has to say that it gives a spine in general because when the triangulated object is a pseudo-manifold any neighborhood of a point does not have the topology of a ball.} of the $3$-manifold represented by the graph \cite{ryan}. 
Furthermore one can define a formal $1/N$ expansion in the case of tensor models. Defining $F[N,\{t_{\mathcal{B}}\}]$ as $Z[N,\{t_{\mathcal{B}}\}]= \exp(-F[N,\{t_{\mathcal{B}}\}])$ the expansion has this form
\be 
F[N,\{t_{\mathcal{B}}\}]=\sum_{\omega \ge 0} N^{D-\frac{2}{(D-1) !}\omega}F_{\omega}[\{t_{\mathcal{B}}\}],
\ee
where we are making some abuse of notations by writing $\omega$ as the sum index instead of the function defined on $(D+1)$-colored graphs taking integer values. 
\subsection{$T^4$ tensor models and intermediate field representation.}

In this section we study the melonic $T^4$ (already studied in \cite{Gurau,Dartois:2013sra}) and write its intermediate field representation using hermitian matrices.

First we introduce some notations. Call $\mathcal{C}$ the set of colors, or equivalently the set that labels the positions of the indices of the components of the tensors. Since a $D$-dimensional tensor model is defined by the use of rank $D$ tensor, $|\mathcal{C}|=D$. For instance, for a tensor of rank three $\mathcal{C}=\{1,2,3\}$, each element of $\mathcal{C}$ indexes respectively the first, the second and the third index of the tensor. 
Moreover we introduce a partial Hermitian product notation, consider a subset of $\mathcal{D} \subset \mathcal{C}$, we denote $\bar T \cdot_{\mathcal{D}} T$ the contraction of all the indices indexed by elements in $\mathcal{D}$, $\bar T \cdot T$ denotes the contraction of all indices. And we denote by $\hat i$ the set $\mathcal{C}-\{i\}$.

In $3$ dimensions using this notations one writes the quartic melonic interaction terms as:
\bea
V[\bar T, T]= \sum_{a=1}^{D=3} (\bar T \cdot_{\hat{a}} T) \cdot_a (\bar T \cdot_{\hat{a}} T)& \nonumber \\
=\sum_{\text{all index}} \bar T_{sjk} T_{s'jk} \bar T_{s'j'k'} T_{sj'k'} + &\bar T_{isk} T_{is'k} \bar T_{i's'k'} T_{i'sk'} + \bar T_{ijs} T_{ijs'} \bar T_{i'j's'} T_{i'j's}
\eea 
By using this notation scheme we write the partition function of the quartic melonic tensor models in $D$-dimensions as:
\begin{eqnarray} \label{t4model}
Z[\lambda, N] = \int dT d\bar T \exp\left[N^{D-1} (-\frac{1}{2}(\bar T \cdot T) - \frac{\lambda}{4}\sum_{a=1}^D (\bar T \cdot_{\hat{a}} T) \cdot_a (\bar T \cdot_{\hat{a}} T))\right].
\end{eqnarray}  
It can be rewritten using intermediate field as: 
\begin{eqnarray}
Z [ \lambda, N ] &=& \int \prod_k d\sigma_k d \bar{\sigma_k}
 \exp\biggl[ - \frac{N}{2} \sum_{c=1\cdots D}   
\Tr(\sigma_c^2)\nonumber \\&&-\Tr\log \left( \mathbbm{1}^{\otimes D}+ i\sqrt{\frac{\lambda}{2N^{D-2}}} \sum_{c=1..D} \mathbbm{1}^{\otimes (c-1)}\otimes \sigma_c \otimes \mathbbm{1}^{\otimes (D-c)}\right) \biggr]   
\end{eqnarray}
One obtains this representation by writing the $T^4$ interaction term as:
\bea \label{above}
&\exp&\left[-N^{D-1} \frac{\lambda}{4} (\bar T \cdot_{\hat{a}} T) \cdot_a (\bar T \cdot_{\hat{a}} T)))\right]\nonumber \\ &=&\int d\sigma_a d\bar \sigma_a \exp\left[-\frac{N}{2} \Tr(\sigma_a^2) - i\sqrt{N^D \lambda/2} \Tr(\Theta^a \sigma_a)\right],
\eea
where $\Theta^a$ denotes $\bar T \cdot_{\hat{a}} T$. Then by replacing the interaction terms of \eqref{t4model} by the right hand side of eq.\eqref{above}, we get:
\bea
Z[\lambda, N] = \int &dT d\bar T& \prod_a d\sigma_a d\bar \sigma_a \exp\biggl[N^{D-1} (-\frac{1}{2}(\bar T \cdot T ))\nonumber \\ &-&\frac{N}{2} \sum_a \Tr(\sigma_a^2) - i\sqrt{N^D \lambda/2} \Tr(\Theta^a \sigma_a)\biggr].
\eea
Integrating out the tensor $\bar T, T$ fields we end with:
\bea \label{Intrep}
Z[\lambda, N] = \int \prod_a d\bar \sigma_a d\sigma_a 
\det\left[\mathbbm{1}^{\otimes D}+ i\sqrt{\frac{\lambda}{2N^{D-2}}}  \sum_{c=1..D} \mathbbm{1}^{\otimes (c-1)}\otimes \sigma_c \otimes \mathbbm{1}^{\otimes (D-c)}\right]^{-1} \nonumber \\ \exp\left[-\frac{N}{2} \sum_a \Tr(\sigma_a^2)\right].
\eea
And so one obtains eq. \eqref{Intrep} by writing: 
\bea
&\det&\left[\mathbbm{1}^{\otimes D}+ i\sqrt{\frac{\lambda}{2N^{D-2}}}  \sum_{c=1..D} \mathbbm{1}^{\otimes (c-1)}\otimes \sigma_c \otimes \mathbbm{1}^{\otimes (D-c)}\right] \nonumber \\ &=&\exp\left[ \Tr \log\left( \mathbbm{1}^{\otimes D}+ i\sqrt{\frac{\lambda}{2N^{D-2}}} \sum_{c=1..D} \mathbbm{1}^{\otimes (c-1)}\otimes \sigma_c \otimes \mathbbm{1}^{\otimes (D-c)}\right)\right]\eea
One notices that expanding the logarithm of \eqref{Intrep} the matrix model one obtains looks like the matrix models introduced in \cite{Borot} for specific choices of the values of the formal variables. The models introduced therein are studied because of their connection to LMO invariants of $3$-manifolds (see \cite{GarMar}). The study of possible connections to tensor models could be interesting and the subject of further works.

\section{Constructing $T^4$ tensor model out of matrix models.} \label{sec:Decomp}
In this section we construct the $T^4$ tensor models as an action of  differential operator acting on a product of hermitian 1-matrix model.
Starting from the intermediate field representation we show:

\begin{theorem}
 The partition function of the $D$-dimensional melonic $T^4$ model can be rewritten as:
 
 \bea
  Z[\lambda, N] =e^{\hat{X}}e^{\hat{Y}} \prod_{i=1}^D Z_{1MM}[\{t^i_p\}_{p\in \mathbb{N}}] = e^{\hat{\mathcal{O}}} \prod_{i=1}^D Z_{1MM}^i[\{t^i_p\}_{p\in \mathbb{N}}]
 \eea
 where $Z_{1MM}[\{t^i_p\}_{p\in \mathbb{N}}]$ is a hermitian 1-matrix model partition function and  $\hat{X}, \hat{Y}, \hat{\mathcal{O}}$ are differential operator acting on the times $t^i_p$ (or coupling constants) of the 1-matrix model: 
 \bea
\hat{X}&=& -\sum_{i,p} t^i_p\frac{\partial}{\partial t^i_p} \\
\hat{Y}&=& \frac{(-1)^D}{N^D}\sum_{(q_1,...,q_D) \in (\mathbb{N}^D)^*} \frac{(-i)^{\sum q_i}}{\sum q_i}\sqrt{\frac{\lambda}{2N^{D-2}}}^{q_1+...+q_D}  \binom{q_1+...+q_D}{q_1,...,q_D} \frac{\partial^D}{\partial t^1_{q_1}...\partial t^D_{q_D}},
 \eea 
 and
  \bea
 \hat{\mathcal{O}} = \ln(e^{\hat{X}}e^{\hat{Y}})= \hat{X}+\frac{D}{2}\frac{\exp(D/2)}{\sinh(D/2)} \hat{Y},
 \eea 
 
\end{theorem}

\textbf{Proof:}

In order to prove this, we make use of the intermediate field representation of the $T^4$ tensor model.

\bea
Z_{T^4_m}[\lambda, N] = \int \prod_k d\sigma_k d \bar{\sigma_k}& \nonumber\\
 \exp\biggl[ - \frac{N}{2} \sum_{c=1..D}   
\Tr(\sigma_c^2)-\Tr&\ln \left( \mathbbm{1}^{\otimes D}+ i\sqrt{\frac{\lambda}{N^{D-2}}} \sum_{c=1..D} \mathbbm{1}^{\otimes (c-1)}\otimes \sigma_c \otimes \mathbbm{1}^{\otimes (D-c)}\right) \biggr] 
\eea
Taylor expanding the logarithmic potential in $\sqrt{\lambda}$ we get:
\bea
Z_{T^4_m}[\lambda, N] = \int \prod_k d\sigma_k d \bar{\sigma_k} 
 \exp\Biggl[ - \frac{N}{2} \sum_{c=1..D}   
\Tr(\sigma_c^2)\Biggr]& \nonumber\\  \exp\Biggl[\sum_{p>0} -\Tr\biggl(  \frac{(-i)^p}{p}\sqrt{\frac{\lambda}{2N^{D-2}}}^p \Bigl(\sum_{c=1..D} & \mathbbm{1}^{\otimes (c-1)}\otimes \sigma_c \otimes \mathbbm{1}^{\otimes (D-c)}\Bigr)^p \biggr)\Biggr] 
\eea
Using multinomial coefficients we can expand $\Tr \Bigl(\sum_{c=1..D}  \mathbbm{1}^{\otimes (c-1)}\otimes \sigma_c \otimes \mathbbm{1}^{\otimes (D-c)}\Bigr)^p$, and so we obtain:
\bea
Z_{T^4_m}[\lambda, N]&=& \int \prod_k d\sigma_k d \bar{\sigma_k} 
 \exp\Biggl[ - \frac{1}{2} \sum_{c=1..D}   
\Tr(\sigma_c^2)\Biggr] \nonumber \\ && \exp\Biggl[\sum_{(q_1,...,q_D)\in(\mathbb{N}^D)^*} \frac{(-i)^{\sum q_i}}{\sum q_i} \sqrt{\frac{\lambda}{2N^{D-2}}}^{\sum q_i} \binom{\sum_i q_i}{q_1,...,q_D} \prod_{c=1}^{D}\Tr(\sigma_c^{q_c})\Biggr] 
\eea
Noticing that for a generic Hermitian 1-matrix model we have the identity:
\bea \label{eq:fundid}
\frac{\partial}{\partial t_p} Z_{1MM}[\{t_p\}] = - N\langle \Tr(\sigma^p) \rangle_{\text{int}}
\eea
We can represent the $T^4$ partition function since in fact:
\bea
Z_{T^4_m}[\lambda, N]&=& \int \prod_k d\sigma_k d \bar{\sigma_k} 
 \exp\Biggl[ - \frac{1}{2} \sum_{c=1..D}   
\Tr(\sigma_c^2)\Biggr] \nonumber \\ &&\sum_{n\ge 0}\frac{1}{n !}\Biggl[\sum_{(q_1,...,q_D)\in(\mathbb{N}^D)^*} \frac{(-i)^{\sum q_i}}{\sum q_i} \sqrt{\frac{\lambda}{2N^{D-2}}}^{\sum q_i} \binom{\sum_i q_i}{q_1,...,q_D} \prod_{c=1}^D \Tr(\sigma_c^{q_c})\Biggr]^n \nonumber \\ &=&
\Biggl\langle \sum_{n\ge 0}\frac{1}{n !}\Biggl[\sum_{(q_1,...,q_D)\in(\mathbb{N}^D)^*} \frac{(-i)^{\sum q_i}}{\sum q_i} \sqrt{\frac{\lambda}{2N^{D-2}}}^{\sum q_i} \binom{\sum_i q_i}{q_1,\cdots,q_D}\prod_{c=1}^D \Tr(\sigma_c^{q_c}))\Biggr]^n \Biggr\rangle_{\text{gau$\beta$}}
\eea
Using equation \ref{eq:fundid} one can write similar correlation functions by differentiating product of matrix models:
\bea
\biggl \langle \frac{(-i)^{\sum q_i}}{\sum q_i} \sqrt{\frac{\lambda}{2N^{D-2}}}^{\sum q_i} \binom{\sum_i q_i}{q_1,...,q_D}\prod_{c=1}^D\Tr(\sigma_c^{q_c}) \biggl \rangle_{\text{int}} & \nonumber \\ 
= \frac{(-1)^D(-i)^{\sum q_i}}{N^D\sum q_i} \sqrt{\frac{\lambda}{2N^{D-2}}}^{\sum q_i} \binom{\sum_i q_i}{q_1,...,q_D} \frac{\partial^D}{\partial t^1_{q_1} \partial t^2_{q_2} ... \partial t^D_{q_D}}& \prod_{i=1}^D
Z_{1MM}[\{t^i_p\}_{p=0}^{\infty}].
\eea
It follows: 
\bea
&\Biggl\langle \sum_{n\ge 0}\frac{1}{n !}\Biggl[\sum_{(q_1,...,q_D)\in(\mathbb{N}^D)^*} \frac{(-i)^{\sum q_i}}{\sum q_i} \sqrt{\frac{\lambda}{2N^{D-2}}}^{\sum q_i} \binom{\sum_i q_i}{q_1,\cdots,q_D} \prod_{c=1}^D\Tr(\sigma_c^{q_c})\Biggr]^n \Biggr\rangle_{\text{int}} \nonumber \\
=&\exp\Biggl[\sum_{(q_1,...,q_D)\in(\mathbb{N}^D)^*} \frac{(-1)^D(-i)^{\sum q_i}}{N^D\sum q_i} \sqrt{\frac{\lambda}{2N^{D-2}}}^{\sum q_i} \binom{\sum_i q_i}{q_1,...,q_D} \frac{\partial^D}{\partial t^1_{q_1} \partial t^2_{q_2} ... \partial t^D_{q_D}} \Biggr] \prod_{i=1}^D
Z^i[\{t^i_p\}_{p=0}^{\infty}], \nonumber \\
\eea
since the differential operators commute. One defines $\hat{Y}$ by:
\be
\hat{Y} =  \sum_{(q_1,...,q_D)\in(\mathbb{N}^D)^*} \frac{(-1)^D(-i)^{\sum q_i}}{N^D\sum q_i} \sqrt{\frac{\lambda}{2N^{D-2}}}^{\sum q_i} \binom{\sum_i q_i}{q_1,...,q_D} \frac{\partial^D}{\partial t^1_{q_1} \partial t^2_{q_2} ... \partial t^D_{q_D}}.
\ee
In order to obtain the gaussian expectation values, and not the interacting one, we act with another operator whose role is to suppress the original matrix potential. In this manner we get the $T^4$ partition function. Define $\hX$ by:
\be
\hat{X} = \sum_{i=1}^D \sum_{p=0}^{\infty} t^{i}_p \frac{\partial}{\partial t^i_p}.
\ee
acting with $\exp({\hat{X}})$ on $\exp(\hat{Y}) \prod_i Z_{1MM}[\{t^i_p\}_{p=0}^{\infty}]$ supresses the matrix potential term as shown by a direct computation.
One wants to find an explicit form for $\hat{\mathcal{O}}$, the commutator of $\hat{X}$ and $\hat{Y}$ is given by:
\bea
[\hat{X},\hat{Y}] = \sum_{(q_1,...,q_D)\in(\mathbb{N}^D)^*} \frac{(-1)^D(-i)^{\sum q_i}}{N^D\sum q_i} \sqrt{\frac{\lambda}{2N^{D-2}}}^{\sum q_i} \binom{\sum_i q_i}{q_1,...,q_D} \frac{\partial^D}{\partial t^1_{q_1} \partial t^2_{q_2} ... \partial t^D_{q_D}} \sum_{i=1}^D \sum_{p=0}^{\infty} t^{i}_p \frac{\partial}{\partial t^i_p} \nonumber \\
= \sum_{(q_1,...,q_D)\in(\mathbb{N}^D)^*} \sum_{i=1}^D \sum_{p=0}^{\infty} \frac{(-1)^D(-i)^{\sum q_i}}{N^D\sum q_i} \sqrt{\frac{\lambda}{2N^{D-2}}}^{\sum q_i} \binom{\sum_i q_i}{q_1,...,q_D} \sum_{j} \delta^{ji} \delta_{pq_j} \frac{\partial^D}{\partial t^1_{q_1} \partial t^2_{q_2} ... \partial t^D_{q_D}} \nonumber \\
= \sum_{(q_1,...,q_D)\in(\mathbb{N}^D)^*} \sum_{i=1}^D \frac{(-1)^D(-i)^{\sum q_i}}{N^D\sum q_i} \sqrt{\frac{\lambda}{2N^{D-2}}}^{\sum q_i} \binom{\sum_i q_i}{q_1,...,q_D} \frac{\partial^D}{\partial t^1_{q_1} \partial t^2_{q_2} ... \partial t^D_{q_D}} \nonumber \\
= D \sum_{(q_1,...,q_D)\in(\mathbb{N}^D)^*} \frac{(-1)^D(-i)^{\sum q_i}}{N^D\sum q_i} \sqrt{\frac{\lambda}{2N^{D-2}}}^{\sum q_i} \binom{\sum_i q_i}{q_1,...,q_D} \frac{\partial^D}{\partial t^1_{q_1} \partial t^2_{q_2} ... \partial t^D_{q_D}}= D \hat{Y}
\eea 
Therefore using Hausdorff-Baker-Campbell formula we can find an explicit form for the operator  $\hat{\mathcal{O}}$:
\bea
\hat{\mathcal{O}} = \log[e^{\hat{X}}e^{\hat{Y}}]= \hat{X} +\frac{D}{1-\exp(-D)} \hat{Y}=\hat{X}+\frac{D}{2} \frac{\exp(D/2)}{\sinh(D/2)} \hat{Y}.
\eea
\qed \\

Paying more attention to the operators $\hat{X}$ and $\hat{Y}$ one notices that they span an $\text{Aff}(1)$ Lie algebra.

\section{Bilinear identities for $T^4$ tensor model.} \label{sec:bilin}

We begin by introducing the orthogonal polynomial for the 1-Hermitian matrix model, we follow closely the presentation given by \cite{Alfaro:1996is}. For each value of the coulpling constant 
$\vec{t}=(t_p)_{p=0\cdots \infty}$  and of $N$ size of the matrix we define:

\begin{definition}
The family of orthogonal polynomials parametrized by $N$ and $\{t_i\}$ in the variable $x$ for the matrix measure is defined by:
\be
P_{N,\vec{t}}=\langle \det(x-M)\rangle_{N,\vec{t}},
\ee
\textit{i.e.} the mean value of the characteristic polynomial of the matrix.
\end{definition}
These polynomials are orthogonal to the matrix measure defined by the partition function of the $1$-Hermitian matrix model. This is only dependant on the fact that the measure comes with a Vandermonde determinant when written in eigenvalues coordinates and that the interaction is symmetric in the coordinates.

For the Hermitian matrix model, the Hirota's equations amount to write the orthogonality relations for the characteristic polynomial. 
They can be written using vertex operators:
\be
\frac{1}{2i\pi}\oint dz \biggl( V_+(z) Z_{1MM}[\{t_i\}]\biggr) \biggl( V_-(z) Z_{1MM}[\{\tilde{t_i}\}]\biggr)=0, 
\ee 
where $V_{\pm}(z)= \exp(\pm \sum_{n\ge 0} z^n t_n) \exp( \mp \log(\frac{1}{z})\frac{\partial}{N\partial t_0} \mp \sum_{n\ge 1} \frac{z^{-n}}{n} \frac{\partial}{N\partial t_n})$.
One would hope the integrable structure of the Hermitian matrix models used for the decomposition survives the action of $\exp(\hat{Y})$. In fact the bilinear Hirota's equations of Hermitian matrix model leads to a set of bilinear identities for the tensor model. Generalizing an idea of \cite{Alfaro:1996is}, one acts by conjugation on the vertex operators of each matrix model of color $c$:
\be
V^c_{\pm}(z, \lambda) = \exp(\hat{Y}) V^c_{\pm}(z) \exp(-\hat{Y}),
\ee
in some sense we make the vertex operators evolves to the intermediate field representation of matrix model \footnote{It's not exactly the tensor model that one obtains, unless one sets the $t_i^c$'s to zero. But one can argue that these bilinear identities are satisfied whatever the value of the $t_i^c$'s and thus induce bilinear identities for the corresponding tensor model.}. And we obtain a set of identities of the form:
\be 
\oint \biggl(V^c_{+}(z, \lambda)\exp(\hat{Y}) \prod_{c'=1}^D Z_{1MM}[\{t_i^{c'} \}]\biggr)\biggl(V^c_{-}(z, \lambda)\exp(\hat{Y})\prod_{c'=1}^D Z_{1MM}[\{\tilde{t}_i^{c'} \}]\biggr)=0,
\ee
for each $c\in [\![ 1,D]\!]$.
Then one has to write these equations in term of the matrices. To this aim we compute the explicit $V^c_{\pm}(z, \lambda)$. First set $\hat{A}^c=\sum_{p\ge 0} z^p t_p^c$ and $\hat{B}^c=\log(\frac{1}{z}) \frac{\partial}{N\partial t^c_0}+\sum_{n=1}^{\infty}\frac{z^{-n}}{n}\frac{\partial}{N\partial t_n^c}$.
Therefore, for $c\in [\![1,D]\!]$:
\bea
&\left[\hat{B}^c, \hY \right] = 0& \nonumber \\
&\left[ \hat{A}^c , \hY \right] =-\frac{(-1)^D}{N^D}&\sum_{(q_1,...,q_D) \in (\mathbb{N}^D)^*} \frac{(-i)^{\sum q_i}}{\sum q_i}\sqrt{\frac{\lambda}{2N^{D-2}}}^{q_1+...+q_D}  \binom{q_1+...+q_D}{q_1,...,q_D}z^{q_c} \frac{\partial^{D-1}}{\partial t^1_{q_1}...\hat{\partial t_{q_c}^c}...\partial t^D_{q_D}}. \nonumber
\eea
One computes the evolved operators explicitly:
\bea
V^c_{\pm}(z, \lambda) &= \exp(\pm \hat{A}^c)\exp(\mp \hat{A}^c)\exp(\hat{Y})\exp(\pm \hat{A}^c) \exp( \mp \hat{B}^c)\exp(-\hat{Y})\nonumber \\
&=\exp(\pm \hat{A}^c)\exp(e^{\mp ad_{\hat{A}^c}}\hat{Y}) \exp( \mp \hat{B}^c)\exp(-\hat{Y}).
\eea
Finally noting that $ad_{\hat{A}^c}^n (Y) = 0$ for $n\ge 2$.
\bea
V^c_{\pm}(z, \lambda)
&=\exp(\pm \hat{A}^c)\exp(\hat{Y}\mp[\hat{A}^c, \hY]) \exp( \mp \hat{B}^c)\exp(-\hat{Y})\nonumber \\
&=\exp(\pm \hat{A}^c)\exp(\mp[\hat{A}^c, \hY]) \exp( \mp \hat{B}^c).
\eea
And so one has proved the following proposition:

\begin{proposition}
The explicit form of the operators $V^c_{\pm}(z, \lambda)$ for $c\in [\![1,D]\!]$ is given by:
\bea
&V^c_{\pm}(z, \lambda)= e^{\pm \sum_{p=0}^{\infty} t^c_p z^p} e^{\mp \log(\frac{1}{z}) \frac{\partial}{N\partial t^c_0}\mp\sum_{n=0}^{\infty}\frac{z^{-n}}{n}\frac{\partial}{N\partial t_n^c}} \nonumber \\
&e^{\pm\frac{(-1)^D}{N^D}\sum_{(q_1,...,q_D) \in (\mathbb{N}^D)^*} \frac{(-i)^{\sum q_i}}{\sum q_i}\sqrt{\frac{\lambda}{2N^{D-2}}}^{q_1+...+q_D}  \binom{q_1+...+q_D}{q_1,...,q_D}z^{q_c} \frac{\partial^{D-1}}{\partial t^1_{q_1}...\hat{\partial t_{q_c}^c}...\partial t^D_{q_D}}}.
\eea
\end{proposition}
Using this proposition we get the form of the Hirota's equation for the intermediate field representation of the tensor model. These can be explicitly rewritten as :
\bea
0&=&\oint dz\; e^{\sum_n z^n(t_n-\tilde{t}_n)} \nonumber \\ 
&&\Biggl\langle\frac{\det\Bigl(z-\sigma_c\Bigr)}{\det\Bigl(\mathbbm{1}^{\otimes D}\bigl(1+z\sqrt{\frac{\lambda}{2N^{D-2}}}\bigr)+\sqrt{\frac{\lambda}{2N^{D-2}}}\sum_{i \neq c}\mathbbm{1}^{\otimes(D-c)}\otimes\sigma_c\otimes \mathbbm{1}^{\otimes(c-1)} \Bigr)}\Biggr\rangle_{N,t} \nonumber \\ 
&&\Biggl\langle\frac{\det\Bigl(\mathbbm{1}^{\otimes D}\bigl(1+z\sqrt{\frac{\lambda}{2N^{D-2}}}\bigr)+\sqrt{\frac{\lambda}{2N^{D-2}}}\sum_{i \neq c}\mathbbm{1}^{\otimes(D-c)}\otimes\sigma_c\otimes \mathbbm{1}^{\otimes(c-1)} \Bigr)}{\det\Bigl(z-\sigma_c\Bigr)}\Biggr\rangle_{N', \tilde{t}}
\eea
We should be able to write these equations in terms of tensor variables. In fact in \cite{VSB} we get a relation between powers of intermediate matrix and power of $\Theta_c$ matrices of the form:
\be
\sigma_c^q= H_q(\Theta^c)
\ee
$H_q$ being the $q^{th}$ Hermite polynomial. We postpone this to future work. 
\section{Conclusion.}

In this paper we unravelled a decomposition of a specific tensor model by the mean of an intertwining operator acting on a product of Hermitian matrix models. The intertwining operator is not of a Givental type since it is not quadratic in the coupling constants, however we do know that the resulting partition function counts specific types of polyangulations of pseudo-manifolds in $D$ dimensions. Moreover this decomposition formula allowed to derive bilinear equations for the tensor model as a 'deformation' of Hirota's equations of the matrix models.

It would be interesting to understand more about this tensor model especially in a more geometric fashion. Can this decomposition help to grasp the geometrical meaning of the number generated by the partition function of the model? Does the observables of this tensor models have anything to do with the symplectic invariants computed by the topological recursion? For some tensor model we know they can be written as matrix model in several way. Do tensor models could provide a framework for writing matrix models that are known to satisfy duality relations between them? A related question being: is there any hope to write Givental models as some tensor models.
At the level of integrability problematic: is the model integrable and can we find any method to compute exactly the model? Answering one or more of these questions could shed light on the real mathematical nature of these tensor models. 

Moreover it should be investigated wether or not it is possibe to generalize this decomposition to arbitrary tensor models in a natural way.

\section*{Acknowledgements.}

I am very grateful to Bertrand Eynard for numerous discussions about Givental decompositions, Topological Recursion and integrability in matrix models. I would also acknowledge Valentin Bonzom for fruitful exchanges about tensor models and integrability therein. I also thank Vincent Rivasseau for continously supporting me during this work. The Institut Galil\'ee and LPT provide excellent working environment. Part of this work has been written at the Erwin Schr\"odinger Institute during the "Combinatorics, Geometry and Physics 2014" program in Vienna. S. Dartois is partially supported by the ANR JCJC CombPhysMat2Tens grant.

 \vskip.5cm
\noindent

{\small ${}^{a}${\it  LIPN, Institut Galil\'ee, CNRS UMR 7030, Universit\'e Paris 13, 
 F-93430, Villetaneuse, France, EU}}
\\
{\small ${}^{b}${\it Laboratoire de Physique Th\'eorique,
Universit\'e Paris 11, 91405 Orsay Cedex, France, EU}} 
\\

\end{document}